# Significantly enhanced polymer thermal conductivity by confining effect through bilayer MoS$_2$ Surfaces

Mohammad Reza Gharib-Zahedi[a*,1], Amin Koochaki[a,b], Mohammad Alaghemandi[c]

[a]Department of Chemistry, Sharif University of Technology, 11365-9516 Tehran, Iran

[b]Department of Chemical Sciences and Bernal Institute, University of Limerick, Limerick V94 T9PX, Republic of Ireland

[c]Department of Electrical and Computer Engineering and BU Photonics Center, Boston University, Boston, MA 02215, USA

## Abstract

The present work refers to a simulation study on nanoconfined polymers in polymer-MoS$_2$ nanocomposites as a function of MoS$_2$-MoS$_2$ interlayer separation. We indeed apply reverse nonequlibrium molecular dynamics simulations (RNEMD) to investigate thermal conductivity of polyamide oligomers which is confined by MoS$_2$ bilayers. The polymer thermal conductivity can be considerably enhanced when polymer chains are confined by MoS$_2$ sheets, in particular such behavior is more obvious for charged surfaces. The presence of the MoS$_2$ surfaces leads to well-ordering of polymer chains as well as denser packing which is an increase in $\lambda$ in the polymer network via the use of MoS$_2$ surface confinement. Polymer chains elongation as well as their preferential alignment parallel to the MoS$_2$ surfaces indicates that $\lambda$ in the polymer domain of the considered nanocomposites is larger than the one in the pure polymer phase. Additionally, our results of number of hydrogen bonds (HBs) in confined polymer chains suggest that a combined effect of the mentioned structural modification and enlarged values of HBs may cooperatively contribute to high polymer thermal conductivity facilitating phonon transport. Results reported here suggest a significant manner to design of confined polymer-MoS$_2$ composites for a wide variety range of applications.

Keywords: Reverse nonequilibrium molecular dynamic, thermal conductivity, confined polymer, MoS$_2$ surfaces

[1)] Correspondence should be addressed: gharibzahedi@gmail.com

# 1. INTRODUCTION

Understanding thermal transport in polymers is essential and is of great significant to a wide range of applications including innovative electronic and optoelectronic [1-5]. However, the low thermal conductivities (λ) in the order of 0.1 W m$^{-1}$ K$^{-1}$ [5,6] hinders the use of polymer materials in these devices, thereby leading to inefficient heat dissipation. Thus, there has been considerable efforts to improve thermal transport properties of polymers [7-12]. To ease the design of nanocomposites with high thermal conductivity, not only is it critical to gain in-depth understanding on the thermal transport in polymeric nanocomposites but also the underlying mechanisms of thermal conductivity in such materials are not completely understood.

Commonly, there is also a competition between existing heat transfer pathways, i.e. collisions and phonons in polymer chains [7-12], and consequently we are always encountered new situation when this competition is altered. It is also the key to determining the thermal conductivity of the polymers. Correspondingly, considerable efforts have been performed to improve thermal conductivity of polymers by e.g. first, introducing high thermal conductivity fillers such as graphene and carbon nanotubes (CNT) into polymer matrix [7,11,12]. Second, changing the morphology of polymer chains including conformational changes by induced anisotropic pressure as well as confinement effect using nano surfaces e.g., metal [13], metal oxides [14], and graphene monolayers [11] as a second. For instances, the high ordered polymer chains is resulted in by graphene surface as relative orientation of the polymer chains in the close of graphene layers, elongated chains compared to pure polymer sample without graphene sheets, and the fluctuation of polymer densities [11]. These observations are more pronounced with increasing graphene surface confinement. Finally, tailoring molecular interactions in order to create new and improve thermal transport pathways in polymer and polymer (nano)composite systems with interchain hydrogen bonding [15-18]. Additionally, since such polymer-CNT (graphene) composites have weak interfaces governed by van der Waals interactions, the thermal conductivity enhancement in these materials is limited, which is significantly below the expectations of the effective medium theory. Therefore, interfacial thermal conductivity may be considerably improved by using non-covalent [19,20] or covalent functionalization [12,21-23] to adjust structural and chemically material interfaces.

Even though some studies concerning λ of polymer matrix by adding nanomaterials have been performed, investigations on the use of monolayer $MoS_2$ possessing excellent optical and electronic properties [24-27] to explore and improve the thermal transport of polymer-$MoS_2$ composites has not been considered up to now. In this work, in an attempt to enhance dramatically the heat transfer performances of polyamide (PA), we have performed systematic calculations of the thermal conductivity of the polymer in a number of polyamide-$MoS_2$ samples by the reverse nonequilibrium molecular dynamics (RNEMD) approach [28,29]. Five different $MoS_2$-$MoS_2$ separations i.e. polymer film thicknesses as interlayer distances were used. The main purpose of the present work is the determination of the thermal conductivities of nanoconfined PA chains between bilayer $MoS_2$ sheets with and without charge. Moreover, their correlation with the structure of the polymer matrix with the aid of different structural properties are analyzed to understand $MoS_2$-prompted changes in the thermal transport mechanisms. The variation of thermal conductivity of polyamide chains as a function of the number of hydrogen bonds is also demonstrated to reveal the relation between thermal conductivity and hydrogen bonding interactions. We consider a pure PA system with the same chain length as met in the polymer-MoS$_2$ composites to enable a comparison with the bulk polymer as well.

## 2. CALCULATION METHOD

We calculate thermal conductivity using the RNEMD simulations [28,29]. The underlying idea of the method is indeed to impose a heat flux as a primary perturbation in the simulation system which is divided to 20 equally thick slabs with adequate particles to guarantee reasonable statistics and to achieve the temperature gradient as a result of the imposed flux. The heat flux is continuously transferred from a "cold" slab, located at the ends of the simulation cell, to the "hot" slab, located at the middle of simulation cell by exchanges of atom velocities. All exchange processes occur under conversation of the total energy between particles with same masses. When the heat flow in the polymer samples reaches the steady state regime (Fig. 1 (c)), the thermal conductivity can be obtained by averaging over the heat flux, temperature gradient, and the Fourier's heat conduction equation (see refs. [28] and [29] for more details).

Therefore, we have limited the velocity exchange processes in the RNEMD approach to the polymer atoms. The MoS$_2$ layers at each end of the simulations' domains have been fixed frozen during the RNEMD processes under NVT ensemble to prevent the heat transfer across the PA chains to the MoS$_2$ sheets, leading all heat flux to cross the PA components. The fixed atoms also prevent the translational dislocation of the whole composites system and as a result we are able to examine a heat transfer restricted to the polymer matrix.

Full atomistic models of nanocomposites were constructed by subjecting polyamide chains to confining between two MoS$_2$ sheets lengthening along the *z* direction, which perpendicular to this direction is also selected as the direction of the heat transfer i.e., x direction parallel to the MoS$_2$ nanosurface. The size of MoS$_2$ sheets is *9 × 3* nm$^2$ (Fig. 1 (a)) and five different confined polymer chains including 15, 36, 64, 90, and 144 PA pentamer chains are studied at 1 atm and temperature 350 K. The atomic structure of PA with five repeating units considering in the study at hand is also illustrated in Fig. 1(b). Polymer-MoS$_2$ systems are indeed confined by placing specific PA chains at a certain interlayer distance of bilayer MoS$_2$ sheets as typically shown in Fig. 1 (d). In order to compare the mentioned systems with polymer pure one, a system more specifically containing 250 PA pentamer chains are simulated. It should be noted that depending to the number of nanoconfined PA chains, the average interlayer distances were approximately in the range of 1 to 9 nm for 15 and 144 PA chains, respectively. It is clear evidence that these conditions will result in solid confinement effects for the polymer chains.

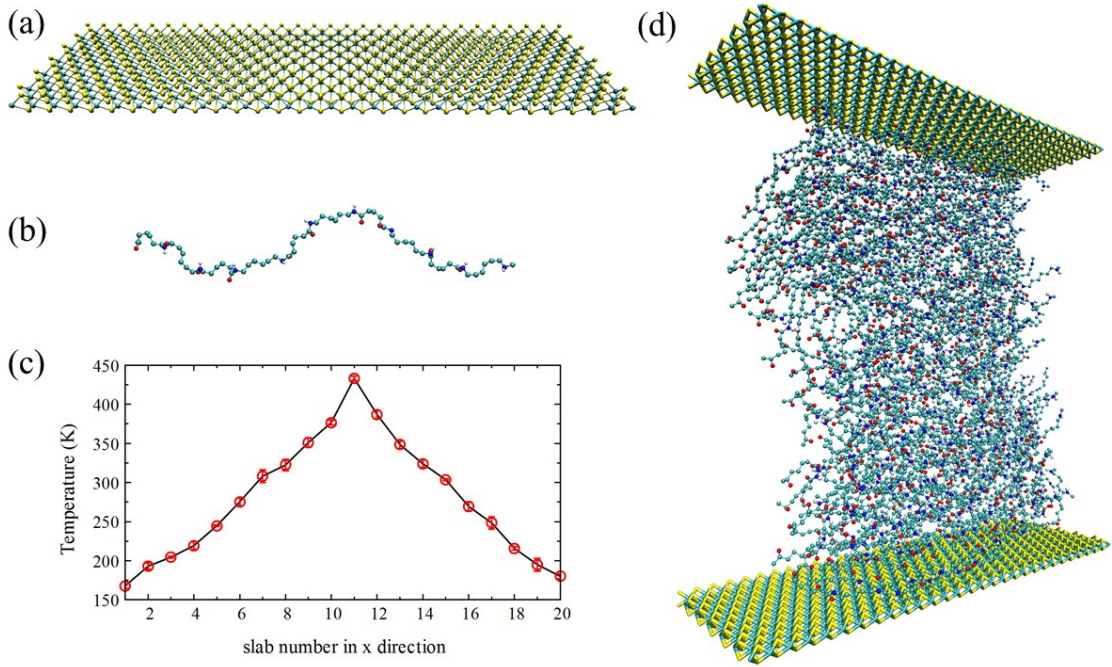

**Figure 1. (a)** MoS$_2$ ($9 \times 3\ nm^2$) and **(b)** atomic structure of PA with five repeating units considered in the present study. **(c)** Typical temperature profile of polymer component as a function of $z$ direction of interlayer distance computed using RNEMD approach. **(d)** The design of the simulated PA-MoS$_2$ nanocomposites including 90 PA polymer chains.

Therefore, in addition to the two mentioned extreme samples, there are three intermediate cases. It should be noted that the interactions between MoS$_2$ sheets and polymer chains can be considered as only with van der Waals interactions, which are modeled by Lennard-Jones potential [30] and also both van der Waals and electrostatic interactions [30]. Because the interactions between MoS$_2$ sheets and PA chains dominate the energy transport across the polymer chains, particularly, in such nanoconfined systems; hence, we adopted the MoS$_2$ sheets with and without charges [30]. To sum up, we studied 10 model systems containing PA chains and MoS$_2$ monolayer sheets in order to derive thermal conductivity of the PA sandwiched in between. The flexible united atom variant (fl-UA) from ref. [31] is used to describe the interaction among PA atoms, and those in MoS$_2$ by a force field derived by Varshney et al. [30]. In this regard, the CH$_2$ and CH$_3$ groups have been considered as uncharged united atoms. In contrast to these uncharged fragments, the amide hydrogen, carbon, nitrogen, and oxygen atoms carried nonzero charges in the adopted fl-UA force field. Moreover, the charge of the Mo and S atoms in MoS$_2$ are 0.734 and -0.367, respectively [30]. The Lorentz-Berthelot mixing rule is adjusted to achieve the Lennard-Jones parameter for the mixed interactions. The nanocomposite systems have been simulated for 25 ns in the NPT statistical ensemble. We have used the Nose-Hoover thermostat and barostat to control the temperature and pressure of the systems with coupling time amount to 0.1 ps and 1.0 ps, respectively. All equilibrium and reverse non-equilibrium MD simulations (RNEMD) were carried out by using large-scale atomic/molecular massively parallel simulator (LAMMPS) [32], with time step of 0.25 fs which is short enough to capture vibrations of light-weighted atoms like hydrogen ones. The velocity-verlet scheme is used to integrate the equation of motion. The non-bonded cutoff is set to 1.2 nm. To evaluate the electrostatic interactions, the Particle-Particle Particle-Mesh (PPPM) method was used with an effective dielectric constant of 5.5 [31].

## 3. RESULTS AND DISCUSSION

The experimental values of $\lambda$ for different polyamide chains have been shown representative inconsistencies 0.15 to 0.30 $W/mK$ [7,12,23]. To validate the employed force field for thermal conductivity simulations, we first calculate the thermal conductivity of pure PA chains. Our simulation setup leads to 0.26 $W/mK$ and 0.29 $W/mK$ of the thermal conductivity of pure PA chains at temperature T of 350 K and 400 K, respectively. Therefore, these values are in line with experimental observations. Moreover, we apply such calculations for a PA chains with ten repeat units and the PA thermal conductivity at T = 350 K has been derived 0.31 $W/mK$ almost similar to pure PA chains with five repeat units. As the results within their error bars show no chain length dependence for polymer thermal conductivity, and thus owing to saving simulation time the polymer system with five repeat units is selected.

Figure 2 shows normalized thermal conductivity ($\lambda_{II}/\lambda_0$) in the polymer component of the studied polyamide–MoS$_2$ composites with varying interlayer distances using RNEMD simulations at 350 K. The $\lambda_{II}$ indicates the PA thermal conductivity in the model systems parallel to the long (z) axis normalizing to the pure PA value, $\lambda_0$. When comparing $\lambda_{II}$ with $\lambda_0$, we observe that the thermal conductivity of polymer all confining between two MoS$_2$ layers highly surpasses the one of bulk polymer. As shown in Figure 2, the highest thermal conductivity is about 35 times larger than the bulk one, which is in the smallest sample with strong confinement effect. In fact, we calculated $\lambda$ of polymer of approximately 9 W. m$^{-1}$.K$^{-1}$ at 350 K for PA polymer chains with highest confinement effect, indicating polymer thermal conductivity enhanced by nearly two orders of magnitude via the use of MoS$_2$ surface confinement. Moreover, a considerable difference between the composites with charged and uncharged MoS$_2$ layers is observed, which is particularly in the systems with small interlayer distances.

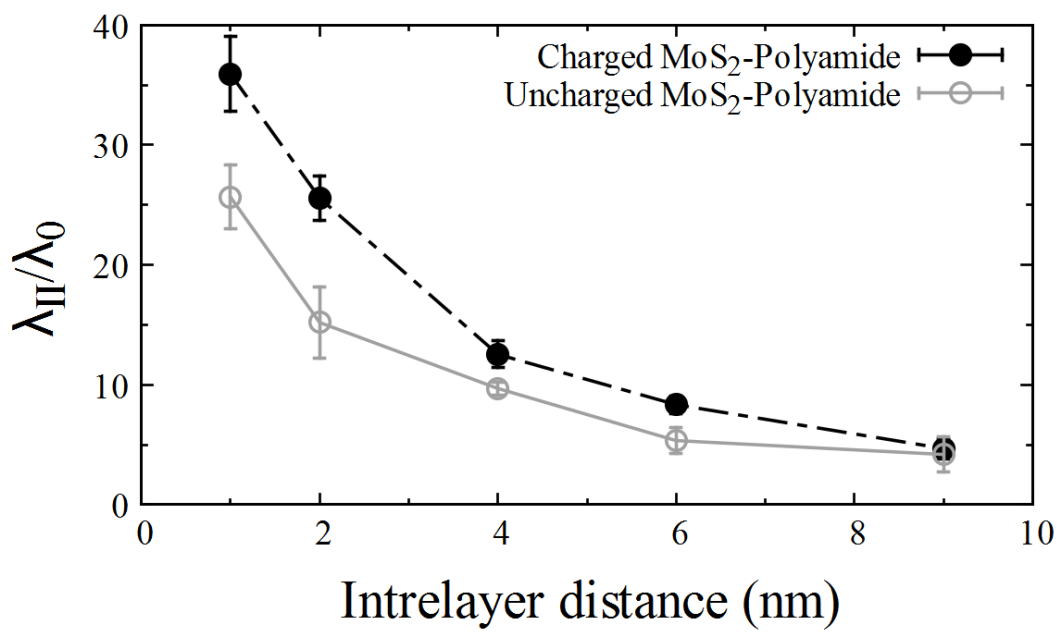

**Figure 2.** Thermal conductivity in the polymer component of the investigated polyamide-MoS$_2$ composites at 350 K as a function of interlayer distance (polymer nanofilm thickness) normalized by the respective bulk value. $\lambda_\parallel$ has been evaluated in the direction of the long x axis (parallel to the Molybdenum disulfide surface). $\lambda_0$ is the thermal conductivity of the pure polymer system at 350 K. The error bars are the standard deviations of the data from the last 2 ns of RNEMD calculation.

To provide the link between the $\lambda$ enhancements to the confinement effect, we investigate structural properties of the polymer in the neighborhood of MoS$_2$ surfaces in the subsequent. Our previous studies showed that such enhancements are due to the formation of the ordered polymer structure in the vicinity of the nanoparticles [7,11,12]. Our investigations indeed found that such ordered structure has a decisive impact on thermal conductivity values. This behavior as well as other key factors responsible for enlarged $\lambda$ in the confined samples will be demonstrated in the following.

Because of paramount importance of the polymer chains density for the properties of the composites, we have firstly analyzed the density of polyamide chains as a function of the distance from the MoS$_2$ layers. Figure 3 presents density profiles for both charged and uncharged MoS$_2$ along the *x* direction in the composite systems studied here. First, the average density profiles show a high peak (maximum) in $\rho(x)$ close to MoS$_2$ surfaces as expected. Although this is not astonishing because of polymer chains are *physically* adsorbed on the surfaces and has been also observed in previous simulations of polymer-solid interfaces with atomistic as well as coarse-grained simulations [34-37], the behavior is more pronounced in the samples with charged MoS$_2$ surfaces, plotted in Figure 3b. It can be concluded that in addition to dispersion van der Waals force between polymer atoms and MoS$_2$ atoms, the strong electrostatic attraction are present between the two-dimensional surface atoms and polyamide ones. Hence, the slight quantitative differences between the density profiles of charged MoS$_2$-polyamide samples related to the value of the first peak reveals the strongest attraction from the MoS$_2$ layers on polymer chains, while the nanocomposites containing uncharged MoS$_2$ feel a weaker attraction. In contrast to all other composites, the 15-chain sample shows a broad maximum for local density. It may be corresponded to decreasing the polymer chains leads to strong confinement effects induced by the MoS$_2$ layers as well as particular MoS$_2$ structure due to existence of positive and negative partial charges together into surface and also inherent geometrical restriction compared to graphene one [11]. Second, beyond the profound peak at a distance equal to more or less 4 Å away from the MoS$_2$ layers, density of PA chains represents a typical fluctuation profiles with considerably lower peaks leading to chain conformations that form a well-ordered layered structure in the vicinity of MoS$_2$ layers. Additionally, away from the surfaces, we observe an established bulk-like PA density and almost symmetric density distribution with respect to the center region of PA films, particularly in samples with large PA chains (90 and 144-chain samples). In all other systems a peak structure is present over the whole interlayer region.

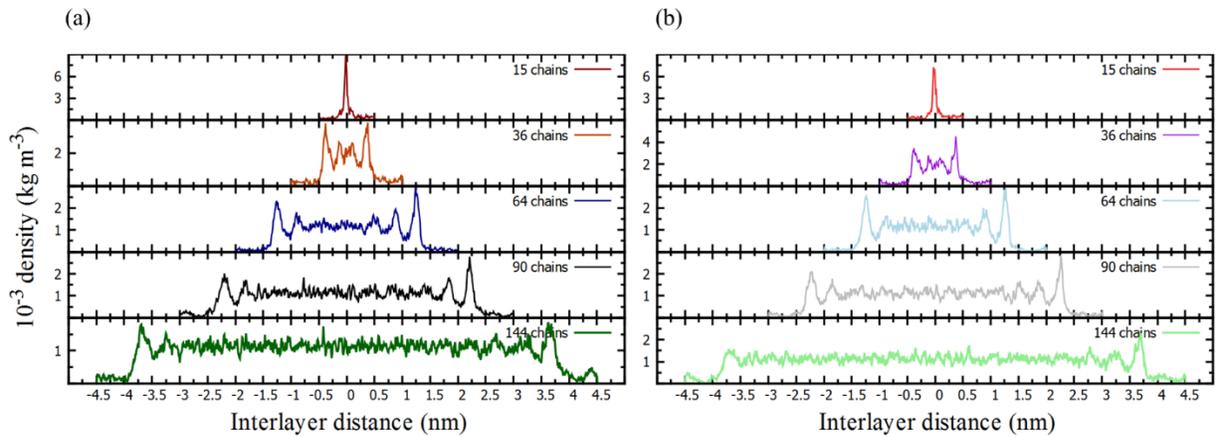

**Figure 3.** Polymer density profile as a function of the interlayer distance (*z*) in the PA-MoS$_2$ systems with uncharged MoS$_2$ surfaces (a) and charged MoS$_2$ surfaces (b).

It is also remarkable to find out how the PA chain conformation is affected by MoS$_2$ surfaces. To determine the effect of the MoS$_2$ confinement on the polymer chains, the variation of the squared radius of gyration $<R_g^2>$ of polymer chains in the nanocomposites can be useful. The $<R_g^2>$ quantities as well as the standard deviations calculated for all polymer chains present in chosen uncharged as well as charged samples are presented in Table 1. We observe that the mean squared radius of gyration for all samples larger than the one of the pure polyamide sample. Additionally, we see that this quantity is enhanced with a decreasing number of PA chains between neighboring MoS$_2$ layers and also for samples with charged surfaces. As a result, the tendency of polymer chains to stretch is enhanced by influencing the surface confinement and charge. It can be concluded that the modifications of $<R_g^2>$ under the influence of the MoS$_2$ surfaces explain enlarged normalized thermal conductivities are shown in Figure 2. Density profiles as well as calculated $<R_g^2>$ for polyamide chains reveal that the confinement made by two dimensional MoS$_2$ surfaces leads to an increased spatial ordering and a significant stretching of the polymer chains in the considered composites in spite of the fact that the pure polymer sample contains an isotropic distribution of the PA bonds.

**Table 1.** Mean-squared radius of gyration $<R_g^2>$ of the polyamide chains in studied nanocomposites with different number of PA chains (different MoS$_2$-MoS$_2$ separation) placed between two MoS$_2$ layers as well as pure polyamide sample. The first numbers are statistical mean values, the second one the standard deviations

| Uncharged MoS$_2$-PA systems | $<R_g^2>$ $(nm^2)$ | Charged MoS$_2$-PA systems | $<R_g^2>$ $(nm^2)$ |
|---|---|---|---|
| 15 chians | 18.19 ± 7.31 | 15 chians | 20.14 ± 8.01 |
| 36 chains | 17.84 ± 8.02 | 36 chains | 18.55 ± 7.83 |
| 44 chains | 16.93 ± 3.22 | 44 chains | 18.21 ± 4.03 |
| 90 chains | 16.04 ± 5.01 | 90 chains | 17.65 ± 4.03 |
| 144 chains | 15.66 ± 3.55 | 144 chains | 17.14 ± 3.43 |
| Pure PA matrix | | | 12.82 ± 2.63 |

To further verify the stretched configuration of polyamide chains in the vicinity of MoS$_2$ surface, the angle $\theta$ between PA bonds and MoS$_2$ sheet (i.e., perpendicular to the interlayer axis) is evaluated. In this respect, the average second-rank order parameter $P_2$ was computed according to $P_2(x) = (\frac{1}{2})(<3\cos^2\theta> -1)$. This parameter spans between three boundaries of -0.5, 0.0 and +1.0 indicating to complete parallel, random, and complete normal orientation situations relative to the MoS$_2$ surface, respectively. The order parameter for C-C bonds in the backbone of PA chains is illustrated only for charged PA-MoS$_2$ nanocomposites in Figure 4 by adopting bins with 2 Å width of the polymer film thickness and the bonds allocating to bins based on the distance of their center-of-mass from the bins. Under the influence of the MoS$_2$ surface, the polymer components close to the surfaces prefer more or less to orient parallel to the layers. Away from the surfaces, we see a slight tendency towards normal orientation relative to the MoS$_2$ surfaces representing a random orientation. The effect of the MoS$_2$ confinement on the orientation preferences of the PA chains in the neighborhood of the surfaces is reduced with increasing count of PA chains *i.e.,* increasing interlayer distances and in fact is more pronounced with decreasing the PA film thickness confined between two layers.

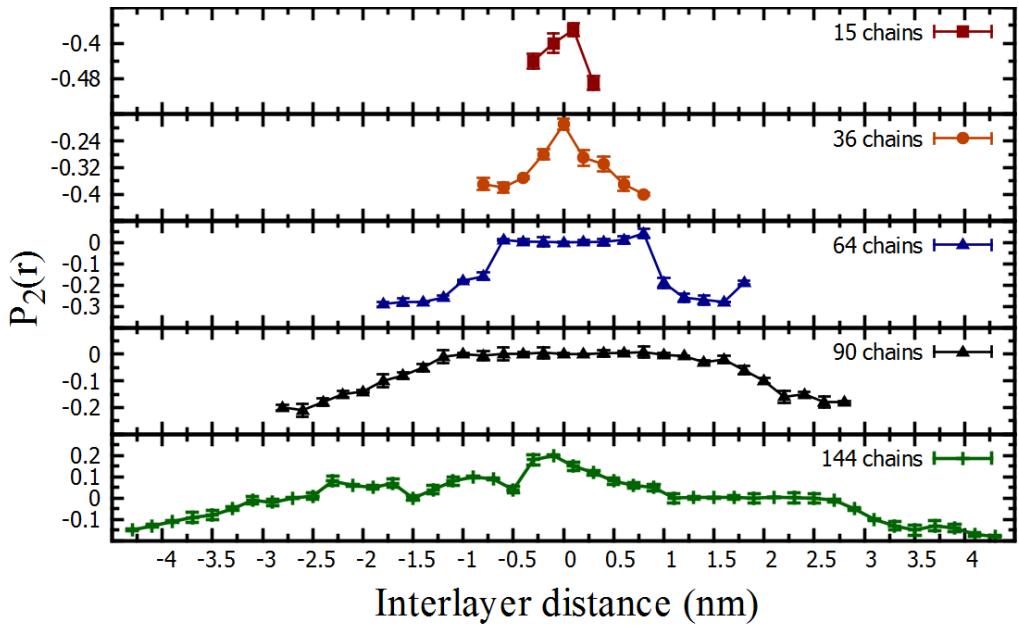

**Figure 4.** Local order parameter for C-C bonds in backbone of polymer chains in the charged composites as a function of MoS$_2$-MoS$_2$ separation.

To further investigate enhanced polymer thermal conductivity in the confined region and its comparison with the pure PA sample, we have explored the role of hydrogen bonding at different interlayer distances. In these systems, two different types hydrogen bonding are expected to be found. Namely, not only the amide hydrogen-amide nitrogen, but also the amide hydrogen-amide oxygen interactions are possible for polyamide chains. The amide (-NH-CO-) as the hydrogen-bond donor amide oxygen as the hydrogen-bond acceptor were chosen. Such selection is because Goudeau et al. [38,39] found that amide nitrogen as an acceptor contributes very small in the hydrogen bonding count for polyamide matrix. The number of hydrogen bonds is calculated with the following geometry criterion, in which the distance between the hydrogen of the donor group, N, and the acceptor O has to be less than 0.297 nm and the donor-hydrogen-acceptor angle has to be bigger than 130° [38-40]. The average number of hydrogen bonds per PA polymer chain to the same quantity in pure PA sample are depicted in Figure 5. As shown in the Figure 5, the degree of hydrogen bonding is strongly influenced by geometrically confined MoS$_2$ surfaces, particularly in the two narrow samples with 16 and 36 PA chains. With increasing distance between the confining surfaces, the number of hydrogen bonds is still higher than the pure PA sample. Selecting a shorter distance as the criterion for hydrogen-bond creation, a similar pattern is detected; however, there are two points to note, one is the ratio of hydrogen bonds in the confined polymers in comparison with the neat polymer system further increased, and the other the amount of increase is higher for nanocomposites samples with interlayer distances 1 and 2 nm. These observations reveal that in confined region the average H···O distance is shorter than that of the pure polymer chain samples. When relating the hydrogen bonding ratio in the Figure 5 to normalized thermal conductivity in Figure 2, we notice that hydrogen bonding can be attributed to increased thermal conductivity of confined polymer chains to assist an efficient thermal transport pathway. This significant effect of hydrogen bonding in confined polymer structures in addition to ordered and elongated polyamide chains as well as preferred aligned chains parallel to nano surfaces lead to an increase in $\lambda$ in the polymer network.

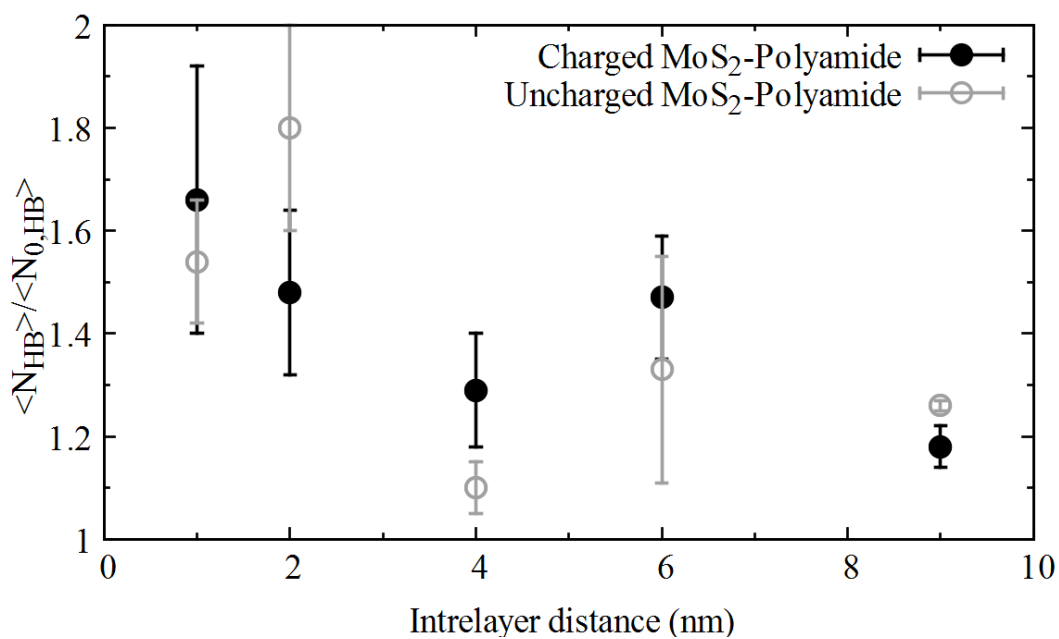

Figure 5. The average number of hydrogen bonds per PA polymer chain to the corresponding quantity for pure PA sample as a function of interlayer distance. Filled and unfilled circles are represents calculated values for composites including charged and uncharged $MoS_2$ layers, respectively.

## 4. CONCLUSION

In conclusion, reverse nonequilibrium molecular dynamics approach was employed to calculate thermal conductivity of nanoconfined polyamide chains between two charged as well as uncharged $MoS_2$ surfaces in a wide range of interlayer separations. Our results show that increased polymer thermal conductivities under the influence of the $MoS_2$ sheets, particularly in highly confined systems. An explanation of these RNEMD results has been possible when adopting structural data of confined polymer chains in the vicinity of the $MoS_2$ bilayer in comparison with the pure polymer sample. The ordering of polymer chains parallel to the surfaces is observed from the number density of PA matrix as a function of separation between the confining surfaces. As a result of electrostatic interactions present in charged $MoS_2$ polymer composites, we can also deduce that the polymer-surface contacts are closer than the ones in the uncharged samples with higher peak intensities. A polymer elongation in the confined polymer chains between $MoS_2$ bilayers is revealed by squared radius of gyration of the PA chains which are larger than the one in the bulk. The stretching of the PA chains is here more pronounced than in the case of the smaller interlayer distance. In addition to these structural changes, it can be seen that the surface effect on the parallel alignment preferences of the PA chains with respect to the surface reduced with increasing polymer thickness i.e. interlayer separation. Moreover, it seems that enlarged values of the average number of hydrogen bonds along with the mentioned structural modifications in nanoconfined polymer matrix contribute supportively in the enhanced $\lambda$ value facilitating phonon transport. Therefore, the calculated thermal conductivity in the polymer domain of the considered nanocomposites is always larger than the one in the pure isotropic polymer phase.